\documentstyle [12pt]{article}
\newcommand{\be}{\begin{equation}}
\newcommand{\ee}{\end{equation}}

\begin{document}
\title{Tetracritical behaviour in a spin-$3/2$ quantum chain}
\author{A. L. Malvezzi\thanks{UFSCARF-TH-94-3. e-mail: palm@power.ufscar.br} \\
	Departamento de F\'\i sica \\ Universidade Federal de S\~ao Carlos \\
	 13565-905, S\~ao Carlos, SP, Brasil}
\date{}
\maketitle
{\centerline{\bf Abstract}}
Finite-Size-Scaling and Conformal Invariance are used in order to find
the phase diagram and critical exponents of a quantum spin chain with
spin $S=3/2$. The model has a tetracritical point besides critical
lines. The conformal anomaly and anomalous dimensions of some primary
operators are calculated at the tetracritical point.
%
%
%
\newpage
Conformal invariance enable us to classify a large number of field theories
in two dimensions \cite{ref1of71,ref3of10}, however it is not obvious that
there exist Statistical Mechanics models realizing such field theories.
Meanwhile severals examples of Statistical Mechanics models described by
such field theories are known \cite{ref4of71}. In particular Andrews, Baxter,
Forrester \cite{ref6of71} and Huse \cite{ref5of71} introduced the RSOS model
which gives a realization of the minimal series \cite{ref3of10} with central
charge
\be
\label{cc}
c = 1 - \frac{6}{(K+2)(K+3)} \hspace{1cm} K = 1,2,3,...\hspace{0.2cm} .
\ee
However the quantum Hamiltonian associated to the RSOS model are rather
complicated.

Recently Sen \cite{ref71}  introduced a series of simple spin-$S$ Hamiltonian.
Based in a Zamolodchikov's Landau-Ginzburg theory he conjectured that, at
special multicritical points, these models also give realizations of the
minimal series in Eq. (\ref{cc}).
These models are defined by the Hamiltonians
\be
\label{H[S]}
H_{[S]} = \sum_{i=1}^{L} \, \left[ \, \frac{1}{2} \left( S_{i}^{z} -
S_{i+1}^{z} \right) + \gamma S_{i}^{x} + \sum_{n=1}^{[S]} \, a_{2n}
\left( S_{i}^{z} \right)^{2n} \right]
\ee
where $S_{i}^{x},S_{i}^{y}$ and $S_{i}^{z}$ are spin-$S$ representations of
$SU(2)$ algebra acting on site $i$ of an $L$ sites chain and periodic boundary
conditions (PBC) are imposed. The integer $K$ in Eq. (\ref{cc}) are related to
the
spin $S$ in Eq. (\ref{H[S]}) by $K=2S$. The degree of the polynomial is $[S]$,
the
largest interger less than or equal $S$. At a particular point such
Hamiltonian is expected \cite{ref71} to present a multicritical point where
$K+1$ phases are indistinguishable. If $S=1/2$ then $K=1$ (\,$c=1/2$\,) and
$H[S]$  [Eq. (\ref{H[S]})] reduces to the Ising Hamiltonian. In the case $S=1$
we have $K=2$ and $H[S]$ is given by
\be
\label{H[3/2]}
H_{[1]} = H_{[3/2]} = -\sum_{i=1}^{L} \, \left[ S_{i}^{z}S_{i+1}^{z}
 - (1+a_{2})\left( S_{i}^{z} \right)^{2} - \gamma S_{i}^{x} \right]
\ee
which is a particular case of the Blume-Emery-Griffths model \cite{ref21of71} ,
having a tricritical point governed by a Conformal Field Theory (CFT) with
$c=7/10$.

In case $S=3/2$ the Hamiltonian is the same as in Eq. (\ref{H[3/2]}) and a
tetracritical point is expected with $c=4/5$ \cite{ref71}. This tetracritical
point is the meeting point where two second order lines end up in a first order
transition line.

In this work by using Finite-Size-Scaling \cite{ref9of72} (FSS) and the
Conformal Invariance (CI) predictions for finite systems
\cite{ref8of10,ref6of10,ref9of10} we studied the Hamiltonian $H[3/2]$
 [Eq. (\ref{H[3/2]})] for $S=3/2$ to
test the conjectures of reference \cite{ref71}.

Firstly, in order to calculate the phase diagram we locate the second order
lines by using FSS. This lines can be found by extrapolating the sequence of
curves obtained from the relation
\be
\label{2cross}
L \, G_{L}(\gamma,a_{2}) = L' \, G_{L'}(\gamma,a_{2})
\ee
where $L$ and $L'$ are the sizes of two finite chains with $L'=L+1$ and
$G_{L}(\gamma,a_{2})$ is the mass gap of the Hamiltonian $H[3/2]$
evaluated at $(\gamma,a_{2})$ for the finite chain of size $L$.

The spectral calculations were done numerically using the Lanczos algorithm
 \cite{ref31oftese} for lattices sizes up to $L=11$. Beyond the translation
invariance the Hamiltonian $H[3/2]$ commutates with the parity operator
\be
\label{parity}
P = \prod_{i=1}^{L} \, \left[ \, \left( S_{i}^{x} \right)^{2} - 1 \, \right]
\ee
and consequently we can separate the Hilbert space into disjoint sectors
labelled according to the momentum $q=\frac{2\pi}{L}l,\, l=0,1,2,...,L-1$
, and parity $p=\pm 1$ eigenvalues.

The ground state (GS) is a zero-momentum state with parity $+\,(-)$ for
lattices
sizes $L$ even (odd), and the first excited state is also a zero-momentum state
but with opposite parity.

In Figure 1 we show in the parameter space of $\gamma$ and $a_{2}$ the
curves satisfying Eq. (\ref{2cross})  for some values of $L$. We also show
 in a large scale the region where the intersections of these curves occur. We
promptly verify the qualitative agreement of Fig. 1 with the phase diagram
proposed in ref. \cite{ref71}.

The tetracritical point P (see Fig.1) where a conformal anomaly $c=4/5$ is
expected \cite{ref71} is obtained by the intersection of the two second
order phase transition lines. Unfortunately, as we can see in Fig. 1, in the
region where these intersections occur, the second order lines are almost
parallel, which make difficult a high precision determination of the
finite-size estimative of the point P. This difficulty obviously will be
reflected in
 the convergence of the numerical sequence used for the bulk limit ($L
\rightarrow \infty$) evaluation of P.

An altenative way \cite{Herrmann}, which give us better estimatives of the
tetracritical point P, is obtained by using the sequence of points calculated
by a generalization of Eq. (\ref{2cross})
\be
\label{3cross}
L \, G_{L}(\gamma,a_{2}) = L' \, G_{L'}(\gamma,a_{2}) =
L'' \, G_{L''}(\gamma,a_{2})
\ee
where $L''=L'+1=L+2$. This condition is satisfyed at the crossing point
obtained
by using Eq. (\ref{2cross}). These crossing points are denoted in Fig. 1 and
we show their coordinates in Table 1. The tetracritical point is obtained by
extrapolating the coordinates of these estimators. Unfortunately, due to the
small number of points these extrapolations are poor, specially for the
sequence related to the coordinate $\gamma$. The best estimatives we get
for these extrapolations are $\gamma = 0.36 \pm 0.01$ and $a_{2}=
0.065 \pm 0.002$, where the errors estimates are personal.

A further analysis of the data of table 1 show us that these points obey,
with a very good fitting, the following relation :
\be
\label{fit}
a_{2}(\gamma) = 0.036883 - 0.23522\gamma - 0.1159\gamma^{2}
\ee
with squares mean deviations 3\,\% and 8\,\% for $a_{2}$ and $\gamma$
respectively. On the other hand in ref. \cite{ref71} based in a
pertubation expansion of $H[3/2]$  up to order $\gamma^{4}$, the point
P should also lay in the curve
\be
\label{serie}
a_{2}(\gamma) = -\frac{1}{2} \gamma^{2} + \frac{25}{192}\gamma^{4}
\hspace{0.2cm} .
\ee
An estimative of P is then obtained by equating Eqs. (\ref{fit}) and
(\ref{serie}). Apart from spurius solutions we obtain P $=(0.3702,-0.0661)$,
which we believe, considering the small number of data we have, to be our
best estimative for this tetracritical point.

Once the point P is calculated the next step towards the verification of the
 conjecture of ref. \cite{ref71} is the calculation of the conformal anomaly
and dimensions of the operators of the underling CFT governing the large
physics at this point. This is done by exploiting at point P the consequences
of CI in the finite-size-corretions of the eigenspectra. The conformal anomaly
 $c$ can be obtained, for periodic chains, from the finite-size-corrections
of the GS energy $E_{0}(L)$, {\em i.e. },
\be
\label{E0ccc}
\frac{E_{0}(L)}{L} = e_{\infty} - \frac{\pi c \zeta}{6L^{2}} + o(L^{-2})
\ee
where $e_{\infty}$ is the bulk limit of the GS energy per particle. In
Eq. (\ref{E0ccc}) $\zeta$ is the sound velocity (non-universal) and can be
calculated from the mass gap associated to the first excitated state with
momentum $\frac{2\pi}{L}$ in the sector with the same parity of the GS,
{\em i.e. },
\be
\label{zetaest}
\zeta_{L} = \frac{L}{2\pi} \, ( E_{0}(L) - E_{1,0} ) \hspace{0.5cm}
 \stackrel{L\rightarrow\infty}{\longrightarrow}
\hspace{0.5cm} \zeta \hspace{0.2cm}.
\ee

Using the Eqs. (\ref{E0ccc}) and (\ref{zetaest}) we obtain for several
points in the second order phase transitions lines of Fig. 1 a value close
to $c=1/2$ (Ising). As we move in Fig. 1 along the second-order transition
line, we verify around the point P a clear cross-over (finite-size-effect)
of our estimatives of the conformal anomaly, and at the point P we obtain
a value $e_{\infty}=-0.2512, \, \zeta=0.67$ and $c=0.76$. Clearly this
is very different of $c=0.5$ of the Ising line and due to numerical
instability in our extrapolations, is also compatible with the conjectured
value \cite{ref71} of $c=0.8$. A better numerical test of the universality
class of the point P is obtained by calculating the anomalous dimensions
of the field theory associated to this point. Associated to the dimension
$x_{\phi}$ of a given primary operator of the theory $\phi$, with spin
$s_{\phi}$, there exist tower of eigenstates with energies and momentum
given by
\be
\label{towerE}
E_{m,m'}(L) - E_{0}(L) = \frac{2\pi\zeta}{L}(x_{\phi} + m + m' )+o(L^{-1})
\ee
\be
\label{towerP}
P_{m,m'}(L) = \frac{2\pi}{L}(s_{\phi} + m - m' ) \hspace{1.0cm} m,m'=
0,1,2,...\,.
\ee
Using these expressions we obtained finite-size estimators for the several
dimensions associated to the eigenenergies of $H[3/2]$. In particular
the lowest gap ( with opposite parity of the GS ) gives the dimension
$x_{\phi_{1}}=0.052$, and in the sector with the parity of the GS the
lowest gap give us $x_{\phi_{2}}=0.130$. These values are much smaller than
the values along the second-order transitions lines ($c=0.5$), which are
$x_{\phi_{1}}=0.125$ and $x_{\phi_{2}}=0.5$. According to the conjecture of
ref. \cite{ref71} these gaps should be compared with the two magnetic operators
of the modular invariant serie~\cite{ref6of10,ref7of10} $A(5)$, with
dimensions $x_{\phi_{1}}=\frac{1}{20}=0.05$ and $x_{\phi_{1}}=
\frac{2}{15}=0.1333...$. As we can see the agreement with our numbers,
taking into account the small number of lattices, is good.

In conclusion, our results for spin $3/2$ and early results for $S=1$ are
clear indications in favour of the conjecture of ref. \cite{ref71}, namely,
the Hamiltonian $H[S]$ [Eq. (\ref{H[S]})] gives us statistical mechanics models
with multicritical points governed by the modular invariance serie $A(K+2)$ of
CFT, having central charge given by Eq. (\ref{cc}) with $K=2S$. After the
completion of our calculations we became aware of a preprint\cite{preprint}
with results in agreement with our conclusions.

It is our pleasure to acknowledge profitable conversations with M. J. Martins
and A. Lima-Santos. We also thanks F. C. Alcaraz to call my atention to
this problem and to Conselho Nacional de Desenvolvimento Cient\'ifico
e Tecnol\'ogico - CNPq - Brasil for financial support
%
%
%

%
%
\newpage
\Large
Figure Captions
\normalsize
\vspace{1.0cm}

Fig. 1 - Curves satisfying  Eq. (\ref{2cross}) for some values of L and,
in a large scale, the region where the intersections of curves occur.
Three lattices crossing points satisfying Eq. (\ref{3cross}) and
tetracritical point P $=(0.3702,-0.0661)$ is also showed.
\vspace{4cm}

\Large
Table Captions
\normalsize
\vspace{1.0cm}

Table 1 - Three lattices crossing points satisfying Eq. (\ref{3cross}) and
extrapolated values.
\newpage
\Large
Table 1
\normalsize
\vspace{1.cm}

\begin{tabular}{||l|l|l||}	\hline
$L,L',L''$	 $\hspace{0.4cm}\gamma$ $\hspace{0.4cm}a_{2}$ \\ \hline
2,3,4		 0.61867	 -0.153 \\
3,4,5		 0.473		 -0.10033 \\
4,5,6		 0.4266	 -0.084526 \\
5,6,7		 0.404		 -0.077047 \\
6,7,8		 0.39166	 -0.073026 \\
7,8,9		 0.384		 -0.070553 \\ \hline
extrap.		 0.36\,$\pm0.01$		 -0.065\,$\pm0.002$ \\ \hline
\end{tabular}
\end{document}